\documentclass[a4paper,12pt,review,oneside,onecolumn]{elsarticle}
\UseRawInputEncoding
\usepackage{times,amsmath,epsfig}
\usepackage{amssymb}
\usepackage{color}
\usepackage{dsfont,epstopdf}
\usepackage{graphicx}
\usepackage{bm}
\usepackage{color}
\usepackage{subfigure} 
\biboptions{numbers,sort&compress}
\usepackage{multirow} 
\usepackage{diagbox}  
\usepackage{makecell} 
\usepackage{geometry}
\newcommand{\ba}{\begin{array}}
\newcommand{\ea}{\end{array}}
\newcommand{\be}{\begin{displaymath}}
\newcommand{\ee}{\end{displaymath}}
\newcommand{\ben}{\begin{equation}}
\newcommand{\een}{\end{equation}}
\newcommand{\bena}{\begin{eqnarray}}
\newcommand{\eena}{\end{eqnarray}}
\newcommand{\beqa}{\begin{eqnarray*}}
\newcommand{\enqa}{\end{eqnarray*}}

\newcommand{\bc}{\begin{center}}
\newcommand{\ec}{\end{center}}
\newcommand{\bi}{\begin{itemize}}
\newcommand{\ei}{\end{itemize}}
\newcommand{\benu}{\begin{enumerate}}
\newcommand{\eenu}{\end{enumerate}}
\newcommand{\bdes}{\begin{description}}
\newcommand{\edes}{\end{description}}
\newcommand{\bt}{\begin{tabular}}
\newcommand{\et}{\end{tabular}}

\newcommand \thetabf{{\mbox{\boldmath$\theta$\unboldmath}}}

\newcommand \alphabf{\mbox{\boldmath$\alpha$\unboldmath}}

\newcommand \Xibf{\hbox{\bf$\Xi$}}

\newcommand \bbf{{\bf b}}

\newcommand \sbf{{\bf s}}

\newcommand \xbf{{\bf x}}

\newcommand \Abf{{\bf A}}
\newcommand \Bbf{{\bf B}}
\newcommand \Cbf{{\bf C}}
\newcommand \Dbf{{\bf D}}
\newcommand \Ebf{{\bf E}}

\newcommand \Ibf{{\bf I}}

\newcommand \Mbf{{\bf M}}
\newcommand \Nbf{{\bf N}}

\newcommand \Pbf{{\bf P}}

\newcommand \Rbf{{\bf R}}
\newcommand \Sbf{{\bf S}}

\newcommand \Ubf{{\bf U}}

\newcommand \Wbf{{\bf W}}
\newcommand \Xbf{{\bf X}}

\newcommand{\circlambda}{\mbox{$\Lambda$
             \kern-.85em\raise1.5ex
             \hbox{$\scriptstyle{\circ}$}}\,}

\renewcommand \thetabf{\boldsymbol{\theta}}

\renewcommand \alphabf{\boldsymbol{\alpha}}

\renewcommand \Xibf{\boldsymbol{\Xi}}

\journal{Signal Processing}

\begin{document}
\begin{frontmatter}
\renewcommand{\thefootnote}{\fnsymbol{footnote}}
\title{Detection of a rank-one signal with limited training data}
\author{Weijian Liu$^{a}$,~Zhaojian Zhang$^{a}$,~Jun Liu$^{b,}$\footnote{Corresponding author. \\    E-mail address: liuvjian@163.com (W. Liu), zzj554038@163.com (Z. Zhang), junliu@ustc.edu.cn (J. Liu), ylwangkjld@163.com (Y.-L. Wang), shangshangzheran@163.com (Z. Shang).}, 
Zheran Shang$^c$,
Yong-Liang Wang$^a$
}
\address{$^a$ Wuhan Electronic Information Institute, Wuhan 430019, China}
\address{$^b$ Department of Electronic Engineering and Information Science, University of Science and Technology of China, Hefei 230027, China} 
\address{$^c$ PLA Academy of Military Science, Beijing 100080, China}
%
%

\begin{abstract}
In this paper, we reconsider the problem of detecting a matrix-valued rank-one signal in unknown Gaussian noise, which was previously addressed for the case of sufficient training data. We relax the above assumption to the case of limited training data. We re-derive the corresponding generalized likelihood ratio test (GLRT) and two-step GLRT (2S--GLRT) based on certain unitary transformation on the test data. It is shown that the re-derived detectors can work with low sample support. Moreover, in sample-abundant environments the re-derived GLRT is the same as the previously proposed GLRT and the re-derived 2S--GLRT has better detection performance than the previously proposed 2S--GLRT. Numerical examples are provided to demonstrate the effectiveness of the re-derived detectors.
\end{abstract}


\end{frontmatter}
\section{Introduction}

Detection of a multichannel signal in unknown noise has received considerable attention from signal processing community 
\cite{DeMaioGreco16Book,DeMaioOrlando16TSP,XueXu18SP_Knowledge,LiuLiu18TSPPHE,ShangLiu18GER, WangZhao19b,LiYang19SP,Besson19,AubryDeMaio20TSP}.
Numerous detectors have been proposed. Precisely, in \cite{Kelly86} Kelly proposed an adaptive detector according to the criterion of generalized likelihood ratio test, referred to as Kelly's GLRT (KGLRT). Then the adaptive matched filter (AMF) was proposed in \cite{RobeyFuhrmann92} based on two-step GLRT. Besides the GLRT, another two common detector design criteria are Rao test and Wald test. For the detection problem in \cite{Kelly86}, the Rao test and Wald test were proposed in \cite{DeMaio07} and \cite{DeMaio04}, respectively, and the Wald test was found to be the same as the AMF. The above detectors are all designed in homogeneous environment, where the training and test data share the same noise covariance matrix. In  \cite{KrautScharf99}, the environment was partially homogeneous in the sense that the training and test data have the same noise covariance matrix upon to an unknown power mismatch, then the adaptive coherence estimator (ACE) was proposed according to the GLRT.
In practice, there often exists signal mismatch. Precisely, the actual signal steering vector is not aligned with the presumed one. Many factors can cause signal mismatch. On the one hand, signal mismatch can be caused by array error, multi-path effect, mutual coupling, etc. In this case, a robust detector is needed \cite{DeMaio05,DeMaioDeNicola09b}. On the other hand, signal mismatch can also be caused by jamming signal entering the radar receiver from sidelobe of the beamformer. In this case, a selective detector is preferred \cite{HaoOrlando14a}. The probability of detection (PD) of a selective detector decreases rapidly with the increase of the amount of signal mismatch. Many selective detectors were proposed in the literature, such as the adaptive beamformer orthogonal rejection test (ABORT) \cite{PulsoneRader01}, whitened ABORT (W-ABORT) \cite{BandieraBesson07TSP_WABORT}, and double-normalized AMF (DN-AMF) \cite{OrlandoRicci10}. For more details about robust and selective detectors, interesting readers can refer to a recent review paper on multichannel adaptive signal detection in \cite{LiuLiu21SCIS}.

The aforementioned detectors were all designed under the assumption of sufficient training data. However, this requirement may not be always satisfied. A kind of method to alleviate the requirement of sufficient is knowledge-aided (KA) method\footnote{KA methods can also be used to solve the problem of adaptive detection or filtering in heterogeneous environment \cite{BessonTourneret07,BidonBesson11AES}.}. One KA method is using existing data, such as historical data, cultural information available from land-use databases, terrain elevation data (DTED) models, etc. \cite{StoicaLiZhu08}. The other KA method is Bayesian model. Precisely, the noise covariance matrices in the test and training data are assumed to be ruled by certain statistical models. Many Bayesian detectors were proposed under this guideline, such as the ones in \cite{DeMaioFarina10,BandieraBesson11}

Most of the above detectors are designed for point targets. However, in practice a target may occupy multiple range bins, especially for high resolution radars.
In \cite{ConteDeMaio01} the problem of distributed target detection is addressed. The echo signals reflected from the distributed target share the same known steering vector but with different unknown signal amplitudes. Several GLRT-based adaptive detectors are designed therein. As mentioned above, in practical applications the signal steering vector may not be exactly known in the case of signal mismatch.
For the distributed target detection, the assumption of the exact knowledge about the spatial steering vector of the signal is relaxed in \cite{BessonScharf06b}. Precisely, it is assumed that the echoes all come from the same direction. However, the spatial steering vector of the signal is only known to lie in a given subspace with unknown coordinates. This problem is tantamount to finding a preferred direction in a known subspace, and hence it is denoted as direction detection in \cite{BessonScharf06b}. According to the 2S--GLRT criterion, the generalized adaptive direction detector (GADD) is proposed in \cite{BessonScharf06b}. In {\cite{LiuLiu2015b} the problem of generalized direction detection is dealt with, which is a generalization of the problem of direction detection. Precisely, the signal to be detected for the problem of generalized direction detection is matrix-valued and both the column and row elements lie in certain known subspaces but with unknown coordinates.

A common assumption in most of the aforementioned detection methods is that sufficient independent and identically distributed (IID) training data are available\footnote{A least requirement for sufficient training data is that the training data can be utilized to form a nonsingular sample covariance matrix (SCM).}.
However, this assumption is usually inappropriate for practical implementations, 
owing to environmental and instrumental factors. To overcome this difficulty,
the \emph{a priori} information about the noise covariance matrix is often utilized, such as low-rank structure \cite{GauReed98,AyoubHaimovich00,JinFriedlander05b}, knowledge-aided information \cite{AubryDeMaio13}, Bayesian methods \cite{DeMaioFarina10,WangSahinoglu11,BandieraBesson11,ZhouZhang12,LiuHan18}, and persymmetric structure \cite{HaoMa12,HaoOrlando14b,GaoLiao14,LiuLiu18TSPPsmtrc,LiuSun19}.
However, the above detectors may suffer from significant performance loss, if the
 \emph{a priori} information imposed on the covariance matrix
 considerably departs from the actual one.

Besides the \emph{a priori} information on the covariance matrix, the signal structure can also be adopted to alleviate the requirement of the training data.
For the problem of generalized direction detection, it is assumed in \cite{BoseSteinhardt96a} that no training data set is available, and no assumption of the \emph{a priori} information on the covariance matrix is adopted. Instead, it is assumed that the test data dimension satisfies a certain constraint.
Precisely, the number of pulses is larger than the sum of the number of the antenna elements and the dimension of the temporal signal subspace.
After a unitary transformation to the test data, a set of the virtual training data set is obtained, which can form an SCM, as an estimate of the unknown covariance matrix. It has been pointed out in \cite{LiuXie14c} that the generalized direction detection problem in \cite{BoseSteinhardt96a} is essentially equivalent to the direction detection with sufficient training data.


In many application, the constraint on data dimension in \cite{BoseSteinhardt96a} may not be hold. Moreover, a more practical case is that there exist training data, but the number is small. This is different from the sufficient-sample assumption in {\cite{LiuLiu2015b} and no-sample assumption in \cite{BoseSteinhardt96a}.
In this paper we reconsider the problem of generalized direction detection in \cite{LiuLiu2015b}. We show that the assumption of sufficient training data is not needed and can be relaxed to a much loose constraint\footnote{Some preliminary results of this paper was given in our previously conference paper in \cite{LiuZhang15}.}. 
To design effective detectors, we first perform a unitary transformation to the test data, which results in a set of virtual training data set. Then, using the virtual and true training data, we re-derive the GLRT and 2S--GLRT. 
A distinct feature of the re-derived detectors is that they can work with low sample support. Moreover, we prove that when the number of the training data is large enough the re-derived GLRT is equivalent to the original GLRT in \cite{LiuLiu2015b}, and the re-derived 2S--GLRT has improved detection performance over the original 2S--GLRT in \cite{LiuLiu2015b}. Moreover, the GLRT and 2S-GLRT in \cite{BoseSteinhardt96a} do not apply under the above loose constraint on data dimension and the number of training data.

The rest of the paper is organized as follows. Section 2 formulates the detection problem, while Section 3 re-derives the GLRT and 2S--GLRT.
Section 4 shows some important properties of the re-derived detectors.
Section 5 presents the simulation results. Finally, some concluding remarks are given in Section 6.

\section{Problem formulation}
The problem of generalized direction detection can be formulated as the following binary hypothesis test \cite{BoseSteinhardt96a,LiuLi19b} \footnote{The subspace model can also be utilized for the case of multiple target detection, with the subspace dimension denoting the number of the targets. However, in this paper we use the subspace model for a single target detection in the case of steering vector uncertainty. The actual signal steering vector is assumed to lie in a carefully-design known subspace but with unknown coordinates.}
\begin{equation}
\label{3}
\left\{ \begin{array}{l}
{\text{H}_0}:\Xbf = \Nbf, \\
{\text{H}_1}:\Xbf = \Abf\bm\theta {\bm\alpha ^H}\Cbf + \Nbf, 
\end{array} \right.
\end{equation}
where
\begin{itemize}
  \item the notation $\text{H}_0$ denotes the signal-absence hypothesis, while $\text{H}_1$ denotes the signal-presence hypothesis, $(\cdot)^H$ denotes the conjugate transpose;
  \item $\Xbf$ is the $N\times K$ test data matrix, $\Nbf$ is the noise matrix in the test data, 
      the columns of $\Nbf$ are mutually independent, subject to complex Gaussian distributions, and share the same but unknown covariance matrix $\Rbf$ \cite{BoseSteinhardt96a,LiuLi19b}.
  \item let $\sbf \triangleq \Abf\thetabf$ and $\bbf\triangleq\Cbf^H\alphabf$, then $\sbf$ is the target's spatial steering vector, and $\bbf$ is the waveform transmitted by the radar system;
  \item $\Abf$ is an $N\times J$ known full-column-rank matrix, whose columns span the subspace where the target's spatial steering vector lies, and the $J\times1$ unknown vector $\thetabf$ stands for the corresponding coordinate;
  \item $\Cbf$ is an $M\times K$ known full-row-rank matrix, whose rows span the subspace where the waveform signal lies, and the $M\times1$ unknown vector $\alphabf$ stands for the corresponding coordinate;
\end{itemize}

The generalized direction detection problem in \eqref{3} can arise for phased-array radars or colocated multiple-input multiple-output (MIMO) radars. Precisely,
for the former the uncertainty in $\bbf$ stands for waveform mismatch \cite{BoseSteinhardt96a}, while for the latter the uncertainty in $\bbf$ stands for transmitting signal steering vector mismatch \cite{LiuLi19b}. Moreover, both for the phased-array radars and colocated MIMO radars, the uncertainty in $\sbf$ stands for receiving signal steering vector mismatch. Waveform mismatch can be caused by polarization diversity or bandpass signals \cite{BoseSteinhardt96a}. Transmitting signal steering vector mismatch and receiving signal steering vector mismatch can both be caused by array errors, multi-path effect, or pointing errors \cite{LiuLi19b,LiuLiu20a}.


It is assumed in \cite{BoseSteinhardt96a} that there are no training data. However, an important constraint is imposed on the data dimension, namely,
\begin{equation}
\label{K_g_mn}
K \ge M + N.
\end{equation}
Based on this constraint, a set of virtual training data can be generated when performing a certain unitary matrix transformation to the test data. These virtual training data are used to form a nonsingular SCM.

It is worth pointing out that training data may be available for some applications, such as the knowledge-aided systems. Precisely, the training data can be land-use and coverage data, or previously collected data over the same terrain \cite{WicksKASTAP}.
In \cite{LiuLiu2015b}, it is assumed  that there are $L$ sufficient training data, denoted as $\xbf_{\text{e},l}$, $l=1,2,\cdots,L$, with $L>N$. These training data can naturally form a nonsingular SCM. 
In fact, the above sufficient-training-data assumption can be relaxed as
\begin{equation}
\label{KL_g_mn}
L + K \ge M+N,
\end{equation}
which is a more practical case; that is there exist training data, but the available training data are limited.

\section{Re-derivations of the GLRT-based detectors under the constraint in \eqref{KL_g_mn}}
To derive the one-step GLRT, we need to maximize the likelihood functions under the two hypotheses in \eqref{3} with training data, and then calculate their ratio.
The joint probability density function (PDF) of the data under hypothesis $\text{H}_1$ is
\begin{equation}
\label{PDF1}
f_1(\Xbf,\Xbf_1)=\frac{\text{e}^{-\left\{\text{tr}(\Rbf^{-1}\Xbf_L\Xbf_L^H)+\text{tr}\left[\Rbf^{-1} (\Xbf-\Abf\thetabf\alphabf^H\Cbf)(\Xbf-\Abf\thetabf\alphabf^H\Cbf)^H\right] \right\}}} {(\pi^N|\Rbf|)^{K+L}},
\end{equation}
where $\Xbf_L=[\xbf_{\text{e},1},\xbf_{\text{e},2},\cdots,\xbf_{\text{e},L}]$ is the training data matrix, the symbols $\text{tr}(\cdot)$ and $|\cdot|$ denote the trace and determinant of a square matrix, respectively.
The maximum likelihood estimate (MLE) of $\Rbf$ under the constraint \eqref{KL_g_mn} is
\begin{equation}
\label{R1}
\hat{\Rbf}_1=\frac{1}{K+L}\left[\Xbf_L\Xbf_L^H+(\Xbf-\Abf\thetabf\alphabf^H\Cbf) (\Xbf-\Abf\thetabf\alphabf^H\Cbf)^H\right],
\end{equation}
which can be written as
\begin{equation}
\label{R1x}
\begin{aligned}
\hat{\Rbf}_1=&\frac{1}{K+L}\left[\Xbf_L\Xbf_L^H+(\Xbf-\Abf\thetabf\alphabf^H\Cbf)\Ubf\Ubf^H (\Xbf-\Abf\thetabf\alphabf^H\Cbf)^H\right]\\
=&\frac{1}{K+L}\left\{\Xbf_L\Xbf_L^H+\left[\Xbf_{_{//}}-\Abf\thetabf\alphabf_\Dbf^H,~\Xbf_ \bot \right] \left[\Xbf_{_{//}}-\Abf\thetabf\alphabf_\Dbf^H,~\Xbf_ \bot \right]^H\right\}\\
=&\frac{1}{K+L}\left[\Sbf_+ +(\Xbf_{_{//}}-\Abf\thetabf\alphabf_\Dbf^H ) (\Xbf_{_{//}}-\Abf\thetabf\alphabf_\Dbf^H )^H\right],
\end{aligned}
\end{equation}
where $\Ubf = [\Cbf_{_{//}}^H,\Cbf_ \bot ^H]$
is a $K \times K$ unitary matrix,
\begin{equation}
\label{Cpp}
\Cbf_{_{//}}=(\Cbf\Cbf^H)^{-\frac{1}{2}}\Cbf,
\end{equation}
${\Cbf_ \bot }$ is a $(K - M) \times K$ semi-unitary matrix and can be obtained by the singular value decomposition (SVD) of $\Cbf$ or $\Cbf_{_{//}}$,
\begin{equation}
\label{7}
{\Xbf_{_{//}}} = \Xbf\Cbf_{_{//}}^H,
\end{equation}
\begin{equation}
\label{8}
{\Xbf_ \bot } = \Xbf\Cbf_ \bot ^H,
\end{equation}
${\bm\alpha _{_\Dbf} } = \Dbf^H\bm\alpha $,  ${\Dbf} = {(\Cbf\Cbf_{}^H)^{\frac{1}{2}}}$,  
\begin{equation}
\label{14}
{\Sbf_ + } = {\Xbf_T}\Xbf_T^H = {\Xbf_L}\Xbf_L^H + {\Xbf_ \bot }\Xbf_ \bot ^H,
\end{equation}
is referred to as the augmented SCM (ASCM), and ${\Xbf_T} = [{\Xbf_ \bot },{\Xbf_L}]$.
The dimensions of ${\Xbf_{_{//}}}$, ${\Xbf_ \bot }$, and ${\Xbf_T}$ are $N \times M$, $N \times (K - M)$, and $N \times (L + K - M)$, respectively.

Note that under hypothesis $\text{H}_1$ the signal component in the test data $\Xbf$ in \eqref{R1x} is ``compressed'' into ${\Xbf_{_{//}}}$ by means of the unitary transformation.
Moreover, it can be verified that the columns of ${\Wbf_{_{//}}}\triangleq \Xbf_{_{//}}-\Abf\thetabf\alphabf_\Dbf^H$ and ${\Xbf_ \bot }$ are IID and share the same covariance matrix $\Rbf$ \cite{LiuZhang15}. 

It is worth pointing out that the augmented matrix ${\Xbf_T} $ can be taken as the new training data set, and the ASCM in \eqref{14} based on the new training data set forms a nonsingular SCM under the constraint in \eqref{KL_g_mn}.
As a result, the problem of generalized direction detection problem in \eqref{3} is converted to the direction detection problem in \cite{LiuXie14c}. Hence, using the results in \cite{LiuXie14c} we can readily obtain the GLRT and 2S--GLRT as
\begin{equation}
\label{12}
\begin{array}{c}
{t_\text{GLRGDD--RU}} = {\lambda _{\max }}\left\{ {\Xbf_{_{//}}^H\Sbf_ + ^{ - 1}\Abf{{({\Abf^H}\Sbf_ + ^{ - 1}\Abf)}^{ - 1}}{\Abf^H}\Sbf_ + ^{ - 1}{\Xbf_{_{//}}}} 
{{{({\Ibf_M} + \Xbf_{_{//}}^H\Sbf_ + ^{ - 1}{\Xbf_{_{//}}})}^{ - 1}}} \right\}
\end{array}
\end{equation}
and
\begin{equation}
\label{13}
{t_\text{AMGDD--RU}} = {\lambda _{\max }}\left\{ {\Xbf_{_{//}}^H\Sbf_ + ^{ - 1}\Abf{{({\Abf^H}\Sbf_ + ^{ - 1}\Abf)}^{ - 1}}{\Abf^H}\Sbf_ + ^{ - 1}{\Xbf_{_{//}}}} \right\},
\end{equation}
respectively, where the notation ${\lambda _{\max }}\{ \cdot \}$ denotes the maximum eigenvalue of a square matrix. 
To differ from the GLRT-based detectors in \cite{LiuLiu2015b}, the detectors in \eqref{12} and \eqref{13} are referred to as the GLR-based generalized direction detector after right unitary transformation (GLRGDD--RU) and adaptive matched generalized direction detector after right unitary transformation (AMGDD--RU)\footnote{The AMGDD--RU is also obtained in \cite{LiuXie14b}. However, it is proposed in \cite{LiuXie14b} in an \emph{ad hoc} manner for a detection problem different from the one in this paper. Precisely, different from the data model in \eqref{3}, it is assumed in \cite{LiuXie14b} that the test data model is $\Xbf = \Abf \Bbf \Cbf + \Nbf$, where the matrix $\Bbf$ is completely unknown and does not have the rank-one structure as shown in \eqref{3}.}, respectively.
A distinct feature of the GLRGDD--RU and AMGDD--RU is that they can work under the condition \eqref{KL_g_mn}, while the original GLRGDD and AMGDD cannot.

The ASCM in \eqref{14} under both hypotheses $\text{H}_0$ and $\text{H}_1$ is ruled by an $N \times N$ complex central Wishart distribution with $L+K-M$ degrees of freedom (DOFs) and a scale matrix $\Rbf$ \cite{Anderson03}. Moreover, the columns of $\Xbf_{_{//}}$ are IID, distributed as complex circular Gaussian distributions, with a common covariance matrix $\Rbf$. Adopting the same approach as in \cite{BoseSteinhardt96a,BandieraBesson07}, one can prove that the GLRGDD--RU and AMGDD--RU have the constant false alarm rate (CFAR) property with respect to the noise covariance matrix $\Rbf$. The proof is omitted for brevity.
Moreover, the computation complexity of the GLRGDD--RU is higher than that of the AMGDD--RU, since the AMGDD--RU needs additional matrix inversion of ${\Ibf_M} + \Xbf_{_{//}}^H\Sbf_ + ^{ - 1}{\Xbf_{_{//}}}$.

A natural question arises: how the GLRGDD--RU and AMGDD--RU behave under the sample-abundant case of $L\geq N$, compared with the original GLRGDD and AMGDD in \cite{LiuLiu2015b}?
This question is answered in the next section.

\section{Comparison of the re-derived detectors with the GLRGDD and AMGDD in \cite{LiuLiu2015b} in the sample-abundant case}
In this section, we 
first show the equivalence of the GLRGDD--RU and GLRGDD in the sample-abundant case of $L \ge N$. 
Next, we show that the AMGDD--RU has better detection performance than the AMGDD when $L \ge N$. By equivalence, we mean that two detectors have the same PDs for a fixed probability of false alarm (PFA).

\subsection{Equivalence of the GLRGDD--RU and GLRGDD When $L \ge N$}
Let $\tilde \Xbf = {\Sbf^{ - \frac{1}{2}}}\Xbf$ and $\tilde \Abf = {\Sbf^{ - \frac{1}{2}}}\Abf$, where ${\Sbf^{ - \frac{1}{2}}} = {({\Sbf^{\frac{1}{2}}})^{ - 1}}$ and ${\Sbf^{\frac{1}{2}}}$ is the squared-root matrix of $\Sbf$. In the sample-abundant case $L \ge N$, the GLRT for the detection problem in \eqref{3}, denoted as the GLRGDD in {\cite{LiuLiu2015b}, is given by
\begin{equation}
\label{16}
\begin{array}{l}
\begin{aligned}
{t_\text{GLRGDD}} = {\lambda _{\max }} \left\{ {\breve{
\Xbf} {\Pbf_{{{\breve{
\Cbf} }^H}}}{{\breve{
\Xbf} }^H}\tilde \Abf} 
{{{ \left [ {{{\tilde \Abf}^H}{{({\Ibf_N} + \tilde \Xbf{{\tilde \Xbf}^H})}^{ - 1}}\tilde \Abf} \right]}^{ - 1}}{{\tilde \Abf}^H}} \right\},
\end{aligned}
\end{array}
\end{equation}
where $\breve{\Xbf}  = \tilde \Xbf{({\Ibf_K} + {\tilde \Xbf^H}\tilde \Xbf)^{ - \frac{1}{2}}}$,
$\breve{\Cbf}  = \Cbf{({\Ibf_K} + {\tilde \Xbf^H}\tilde \Xbf)^{ - \frac{1}{2}}}$,
and
${\Pbf_{{{\breve{\Cbf} }^H}}} = {\breve{\Cbf} ^H}{(\breve{\Cbf} {\breve{\Cbf} ^H})^{ - 1}}\breve{\Cbf}$.

For any two conformable matrices $\Ebf_1$ and $\Ebf_2$, we have
\begin{equation}
\label{29}
\lambda \{\Ebf_1 \Ebf_2\} = \lambda \{\Ebf_2\Ebf_1 \},
\end{equation}
where $\lambda \{ \cdot \}$ stands for an eigenvalue of a square matrix. After some algebra and using \eqref{29}, we can extend \eqref{16} as
\begin{equation}
\label{30}
\begin{array}{r}
\begin{aligned}
{t_\text{GLRGDD}} = {\lambda _{\max }}&\left\{ {{{\left[ {{\Abf^H}{{(\Sbf + \Xbf{\Xbf^H})}^{ - 1}}\Abf} \right]}^{ - 1}}} \right.\\
 &\cdot {\Abf^H}{\Sbf^{ - 1}}\Xbf{({\Ibf_K} + {\Xbf^H}{\Sbf^{ - 1}}\Xbf)^{ - 1}}{\Cbf^H}\\
 &\cdot {\left[ {\Cbf{{({\Ibf_K} + {\Xbf^H}{\Sbf^{ - 1}}\Xbf)}^{ - 1}}{\Cbf^H}} \right]^{ - 1}}\\
 &\cdot \Cbf\left. {{{({\Ibf_K} + {\Xbf^H}{\Sbf^{ - 1}}\Xbf)}^{ - 1}}{\Xbf^H}{\Sbf^{ - 1}}\Abf} \right\}.
\end{aligned}
\end{array}
\end{equation}
Substituting $\Cbf=\Dbf\Cbf_{_{//}}$ into \eqref{30}, with $\Cbf_{_{//}}$ given in \eqref{Cpp}, leads to
\begin{equation}
\label{31}
\begin{array}{l}
\begin{aligned}
{t_\text{GLRGDD}} = {\lambda _{\max }}\left\{ \Xibf_\Abf ~
\Xibf_{\Abf,\Cbf} ~
\Xibf_{\Cbf} ~
\Xibf_{\Abf,\Cbf}^H\right\},
\end{aligned}
\end{array}
\end{equation}
where
\begin{equation}
\label{XiA}
\Xibf_\Abf={{{[ {{\Abf^H}{{(\Sbf + \Xbf{\Xbf^H})}^{ - 1}}\Abf} ]}^{ - 1}}},
\end{equation}
$\Sbf = {\Xbf_L}\Xbf_L^H$
is the conventional SCM,
\begin{equation}
\label{XiAC}
\Xibf_{\Abf,\Cbf}={\Abf^H}{\Sbf^{ - 1}}\Xbf{({\Ibf_K} + {\Xbf^H}{\Sbf^{ - 1}}\Xbf)^{ - 1}}\Cbf_{_{//}}^H,
\end{equation}
and
\begin{equation}
\label{XiC}
\Xibf_{\Cbf}={[ {{\Cbf_{_{//}}}{{({\Ibf_K} + {\Xbf^H}{\Sbf^{ - 1}}\Xbf)}^{ - 1}}\Cbf_{_{//}}^H} ]^{ - 1}}.\end{equation}

It is shown in Appendix A that \eqref{12} is equivalent to \eqref{31} when $L \ge N$. Hence, the GLRGDD--RU and GLRGDD have the same detection performance in the sample-abundant environment. However, it is worth noting that the GLRTDD is invalid with low sample support of $L < N$, due to the singularity of the original SCM $\Sbf$.

\subsection{Superiority of the AMGDD--RU over the AMGDD When $L \ge N$}
When $L \ge N$ the 2S--GLRT for the detection problem in \eqref{3}, referred to as the AMGDD in {\cite{LiuLiu2015b}, is
\begin{equation}
\label{20}
{t_\text{AMGDD}} = {\lambda _{\max }}\{{\Pbf_{{\Cbf^H}}}{\tilde \Xbf^H}{\Pbf_{\tilde \Abf}}\tilde \Xbf{\Pbf_{{\Cbf^H}}}\},
\end{equation}
where ${\Pbf_{{\Cbf^H}}} = {\Cbf^H}{(\Cbf{\Cbf^H})^{ - 1}}\Cbf$
and ${\Pbf_{\tilde \Abf}} = \tilde \Abf{({\tilde \Abf^H}\tilde \Abf)^{ - 1}}{\tilde \Abf^H}$.
Equation \eqref{20} can be rewritten as
\begin{equation}
\label{23}
\begin{array}{c}
\begin{aligned}
{t_\text{AMGDD}} = {\lambda _{\max }}\left\{ {\Pbf_{{\Cbf^H}}}{\Xbf^H}{\Sbf^{ - 1}}\Abf{({\Abf^H}{\Sbf^{ - 1}}\Abf)^{ - 1}} {\Abf^H}{\Sbf^{ - 1}}\Xbf{\Pbf_{{\Cbf^H}}}\right\}.
\end{aligned}
\end{array}
\end{equation}
Substituting $\Cbf=\Dbf\Cbf_{_{//}}$ into 
the definition of ${\Pbf_{{\Cbf^H}}}$,
with $\Cbf_{_{//}}$ given in \eqref{Cpp}, leads to
\begin{equation}
\label{24}
{\Pbf_{{\Cbf^H}}} = \Cbf_{_{//}}^H{\Cbf_{_{//}}},
\end{equation}
Inserting \eqref{24} into \eqref{23} and using \eqref{29}, we arrive at
\begin{equation}
\label{25}
{t_\text{AMGDD}} = {\lambda _{\max }}\{\Xbf_{_{//}}^H{\Sbf^{ - 1}}\Abf{({\Abf^H}{\Sbf^{ - 1}}\Abf)^{ - 1}}{\Abf^H}{\Sbf^{ - 1}}\Xbf_{_{//}}\},
\end{equation}
where $\Xbf_{_{//}}^{}$ is given in \eqref{7}.

Inspection of \eqref{13} and \eqref{25} 
reveals that the AMGDD and AMGDD--RU have similar forms. The main difference is that the former uses the true training data ${\Xbf_L}$ to estimate the unknown covariance matrix \Rbf, while the latter uses both the true training data ${\Xbf_L}$ and the virtual training data ${\Xbf_ \bot }$ given in \eqref{8} to estimate $\Rbf$. Hence, it is expected that the AMGDD--RU has better detection performance than the AMGDD, since the AMGDD--RU employs more training data. This is can be further confirmed by simulation results in Section 5.

Before closing this section, we give some preliminary analysis of computational complexity. The main computational complexity of the proposed GLRDD-RU and AMGDD-RU are the matrix inversion and singular value decomposition. Precisely, the computational complexity for the GLRGDD-RU is $\text{O}({{r}^{3}})$, which is the same as the GLRGDD with $r=\max (N,K)$. This is because the GLRGDD-RU needs to compute the $K\times K$ matrix $\mathbf{U}$, defined below (6), and the inversion of the $N\times N$ matrix ${{\mathbf{S}}_{+}}$. The fact that the computational complexity for the GLRGDD is $\text{O}({{r}^{3}})$ is due to the operation of ${{\text{(}{{\mathbf{I}}_{K}}\text{+}{{\mathbf{X}}^{H}}{{\mathbf{S}}^{-1}}\mathbf{X}\text{)}}^{-1}}$, shown in (15). In a similar manner, it can be shown that the computational complexity for the AMGDD-RU is also $\text{O(}{{r}^{3}}\text{)}$, and the computational complexity for the AMGDD is also $\text{O(}{{p}^{3}}\text{)}$, with $p=\max (N,M)$. In summary, the GLRGDD-RU has the same computational complexity with the GLRGDD, while the AMGDD-RU has the same or higher computational complexity than the AMGDD.

\section{Numerical examples}
In this section we compare the detection performance of the GLRGDD--RU and AMGDD--RU with the GLRGDD in \eqref{16}, AMGDD in \eqref{20}, and Bose's GLRT in \cite{BoseSteinhardt96a}, given by
\begin{equation}
\label{27}
{t_\text{GLRT}} = {\lambda _{\max }}  \left\{ {\Xbf_{_{//}}^H\Sbf_ \bot ^{ - 1}\Abf{{({\Abf^H}\Sbf_ \bot ^{ - 1}\Abf)}^{ - 1}}{\Abf^H}\Sbf_ \bot ^{ - 1}{\Xbf_{_{//}}}} 
{{{({\Ibf_M} + \Xbf_{_{//}}^H\Sbf_ \bot ^{ - 1}{\Xbf_{_{//}}})}^{ - 1}}} \right\},
\end{equation}
which is valid only when $K\ge M+N$. In \eqref{27},
\begin{equation}
\label{15}
{\Sbf_ \bot } = {\Xbf_ \bot }\Xbf_ \bot ^H.
\end{equation}

For Monte Carlo simulations, $10^4$ data realizations are carried out to evaluate the PD.  $10^5$ data realizations are generated to determine the detection threshold, according to a preassigned PFA, which is chosen as $10^{-3}$. The $(i,j)$ element of $\Rbf$ is set to be ${0.95^{|i - j|}}$, with $| \cdot |$ standing for the modulus of a real number. $\Abf$ and $\Cbf$ are randomly chosen. After being generated they are fixed in each Monte Carlo simulation.

Fig. 
1 displays the PDs of the detectors as functions of the signal-to-noise ratio (SNR) in the sample-abundant case of $L \ge N$ and the dimension constraint in \eqref{K_g_mn} holds. The SNR is defined as
\begin{equation}
\label{28}
\text{SNR} = {\bm\alpha ^H}\Cbf{\Cbf^H}\bm\alpha  \cdot {\bm\theta ^H}{\Abf^H}{\Rbf^{ - 1}}\Abf\bm\theta.
\end{equation}
The results indicate that the GLRGDD--RU and AMGDD--RU significantly outperform the AMGDD and Bose's GLRT. 
Moreover, the GLRGDD--RU coincides with the GLRGDD, which is consistent with the theoretical prediction in Section 4.1. In addition, the GLRGDD--RU (or GLRGDD) has a slightly higher PD than the AMGDD--RU. 

Fig. 
2 show the detection performance of the GLRGDD--RU and AMGDD--RU with low sample support of $L < N$ and $K < M + N$. The GLRGDD, AMGDD, and Bose's GLRT are not plotted herein, since they fail to work in this scenario. It is seen that the detection performance of the GLRGDD--RU and AMGDD--RU improves with the increase of $K$. Essentially, this is due to the increase in the number of the virtual training data, as indicated in \eqref{14}. Comparing 
the results in Figs. 
1 and 2, we see that the performance improvement of the GLRGDD--RU to the AMGDD--RU is more obvious with low sample support.


\section{Conclusions}
In this paper we re-derived the GLRT and 2S--GLRT for the generalized direction detection with limited training data by using unitary transformation, resulting in the detectors GLRGDD--RU and AMGDD--RU.
The GLRGDD--RU and AMGDD--RU possess the CFAR property with respect to the unknown noise covariance matrix, and can work with low sample support where the number of the training data is small such that the original SCM is singular, i.e., $L<N$, but satisfies $L+K \ge M+N$. Meanwhile, in the sample-abundant case of $L \ge N$ the GLRGDD--RU is equivalent to the GLRGDD. In addition, the AMGDD--RU achieves much better detection performance than the AMGDD, since the former employs additional virtual training data.

\section*{Acknowledgements}
This work was partially supported by the National Key Research and Development Program of China (No. 2018YFB1801105), the National Natural Science Foundation of China (Nos. 62071482 and 61871469), and the Youth Innovation Promotion Association CAS (No. CX2100060053).

\appendix
\section{Equivalence of \eqref{12} and \eqref{31} in the sample-abundant case of $L\ge N$}

In the following we derive the equivalent forms of  
\eqref{XiA}, and \eqref{XiAC}, and \eqref{XiC}.
Then we use these results to prove the equivalence of the GLRGDD--RU and GLRGDD.

\subsection{ Equivalent form of \eqref{XiA}}
According to the matrix inversion lemma \cite[p.534]{Kay98}, we have
\begin{equation}
\label{32}
\begin{array}{l}
\begin{aligned}
{({\Bbf_0} + {\Bbf_1}{\Bbf_2}{\Bbf_3})^{ - 1}} = \Bbf_0^{ - 1}- \Bbf_0^{ - 1}{\Bbf_1}
{(\Bbf_2^{ - 1} + {\Bbf_3}\Bbf_0^{ - 1}{\Bbf_1})^{ - 1}}{\Bbf_3}\Bbf_0^{ - 1},
\end{aligned}
\end{array}
\end{equation}
where ${\Bbf_i}$’s, $i = 0,1,2,3,$ are arbitrary conformable matrices. Moreover, according to the definitions of ${\Xbf_ \bot } $ and $\Xbf_{_{//}}$, we obtain
\begin{equation}
\label{33}
\Xbf{\Xbf^H} = {\Xbf_ \bot }\Xbf_ \bot ^H + {\Xbf_{_{//}}}\Xbf_{_{//}}^H.
\end{equation}
In view of \eqref{32} and \eqref{33}, it follows
\begin{equation}
\label{34}
\begin{array}{c}
\begin{aligned}
{(\Sbf + \Xbf{\Xbf^H})^{ - 1}}&= {({\Sbf_ + } + {\Xbf_{_{//}}}\Xbf_{_{//}}^H)^{ - 1}}\\
&= \Sbf_ + ^{ - 1}- \Sbf_ + ^{ - 1}{\Xbf_{_{//}}}
 {({\Ibf_M} + \Xbf_{_{//}}^H\Sbf_ + ^{ - 1}{\Xbf_{_{//}}})^{ - 1}}\Xbf_{_{//}}^H\Sbf_ + ^{ - 1},
 \end{aligned}
\end{array}
\end{equation}
where ${\Sbf_ + }$ is given in \eqref{14}. In light of \eqref{34}, we have
\begin{equation}
\label{35}
{\Abf^H}{(\Sbf + \Xbf{\Xbf^H})^{ - 1}}\Abf = {\bm\Phi _\Abf} - {\bm\Phi _{\Abf \Xbf}}{({\Ibf_M} + {\bm\Phi _\Xbf})^{ - 1}}{\bm\Phi _{\Xbf\Abf}},
\end{equation}
where
\begin{equation}
\label{36}
{\bm\Phi _\Abf} = {\Abf^H}\Sbf_ + ^{ - 1}\Abf,
\end{equation}
\begin{equation}
\label{37}
{\bm\Phi _{\Abf\Xbf}} = {\Abf^H}\Sbf_ + ^{ - 1}{\Xbf_{_{//}}},
\end{equation}
\begin{equation}
\label{38}
{\bm\Phi _{\Xbf\Abf}} = \Xbf_{_{//}}^H\Sbf_ + ^{ - 1}\Abf,
\end{equation}
\begin{equation}
\label{39}
{\bm\Phi _\Xbf} = \Xbf_{_{//}}^H\Sbf_ + ^{ - 1}{\Xbf_{_{//}}}.
\end{equation}
Performing the matrix inversion to \eqref{35} and using \eqref{32} leads to
\begin{equation}
\label{40}
\Xibf_\Abf= \bm\Phi _\Abf^{ - 1}+\bm\Phi _\Abf^{ - 1}{\bm\Phi _{\Abf\Xbf}} 
 {({\Ibf_M} + {\bm\Phi _\Xbf} - {\bm\Phi _{\Xbf\Abf}}\bm\Phi _\Abf^{ - 1}{\bm\Phi _{\Abf\Xbf}})^{ - 1}}{\bm\Phi _{\Xbf\Abf}}\bm\Phi _\Abf^{ - 1}.
\end{equation}
\subsection{ Equivalent form of \eqref{XiAC}}
According to \eqref{32}, we have
\begin{equation}
\label{41}
{({\Ibf_K} + {\Xbf^H}{\Sbf^{ - 1}}\Xbf)^{ - 1}} = {\Ibf_K} - {\Xbf^H}{(\Sbf + \Xbf{\Xbf^H})^{ - 1}}\Xbf.
\end{equation}
Substituting \eqref{34} into \eqref{41} results in
\begin{equation}
\label{42}
{({\Ibf_K} + {\Xbf^H}{\Sbf^{ - 1}}\Xbf)^{ - 1}} = {\Ibf_K} - {\Xbf^H}\Sbf_ + ^{ - 1}\Xbf
+ {\Xbf^H}\Sbf_ + ^{ - 1}{\Xbf_{_{//}}}{({\Ibf_M} + {\bm\Phi _\Xbf})^{ - 1}}\Xbf_{_{//}}^H\Sbf_ + ^{ - 1}\Xbf.
\end{equation}
It follows that
\begin{equation}
\label{43}
\begin{array}{l}
\begin{aligned}
\Xibf_{\Abf,\Cbf}=~ &{\Abf^H}{\Sbf^{ - 1}}{\Xbf_{_{//}}}- {\Abf^H}{\Sbf^{ - 1}}\Xbf{\Xbf^H}\Sbf_ + ^{ - 1}{\Xbf_{_{//}}}\\
&+ {\Abf^H}{\Sbf^{ - 1}}\Xbf{\Xbf^H}\Sbf_ + ^{ - 1}{\Xbf_{_{//}}}{({\Ibf_M} + {\bm\Phi _\Xbf})^{ - 1}}{\bm\Phi _\Xbf},
\end{aligned}
\end{array}
\end{equation}
where ${\bm\Phi _\Xbf}$ is given in \eqref{39}. Plugging \eqref{33} into \eqref{43}, after some algebra, yields
\begin{equation}
\label{44}
\begin{array}{l}
\begin{aligned}
\Xibf_{\Abf,\Cbf}=~ &{\Abf^H}{\Sbf^{ - 1}}{\Xbf_{_{//}}} - {\Abf^H}{\Sbf^{ - 1}}{\Xbf_{_{//}}}{\bm\Phi _\Xbf}\\
&- {\Abf^H}{\Sbf^{ - 1}}{\Xbf_ \bot }\Xbf_ \bot ^H\Sbf_ + ^{ - 1}{\Xbf_{_{//}}} \\
&+ {\Abf^H}{\Sbf^{ - 1}}{\Xbf_{_{//}}}{\bm\Phi _\Xbf}{({\Ibf_M} + {\bm\Phi _\Xbf})^{ - 1}}{\bm\Phi _\Xbf}\\
{\kern 1pt} {\kern 1pt} {\kern 1pt} {\kern 1pt} {\kern 1pt} {\kern 1pt} {\kern 1pt} {\kern 1pt} {\kern 1pt} {\kern 1pt} {\kern 1pt} {\kern 1pt} {\kern 1pt} {\kern 1pt} {\kern 1pt} {\kern 1pt}  &+ {\Abf^H}{\Sbf^{ - 1}}{\Xbf_ \bot }\Xbf_ \bot ^H\Sbf_ + ^{ - 1}{\Xbf_{_{//}}}{({\Ibf_M} + {\bm\Phi _\Xbf})^{ - 1}}{\bm\Phi _\Xbf}.
\end{aligned}
\end{array}
\end{equation}
The first, second, and fourth addends in \eqref{44} can be collected as
\begin{equation}
\label{45}
{\Abf^H}{\Sbf^{ - 1}}{\Xbf_{_{//}}}[{\Ibf_M} - {\bm\Phi _\Xbf} + {\bm\Phi _\Xbf}{({\Ibf_M} + {\bm\Phi _\Xbf})^{ - 1}}{\bm\Phi _\Xbf}] = 
{\Abf^H}{\Sbf^{ - 1}}{\Xbf_{_{//}}}{({\Ibf_M} + {\bm\Phi _\Xbf})^{ - 1}},
\end{equation}
where we have used the identity \cite{BandieraBesson07TSP_WABORT}
\begin{equation}
\label{46}
	{\Ibf_M} - {\bm\Phi _\Xbf} + {\bm\Phi _\Xbf}{({\Ibf_M} + {\bm\Phi _\Xbf})^{ - 1}}{\bm\Phi _\Xbf} = {({\Ibf_M} + {\bm\Phi _\Xbf})^{ - 1}}.
\end{equation}
The third and fifth addends in \eqref{44} can be rewritten as
\begin{equation}
\label{47}
	\begin{array}{l}
 - {\Abf^H}{\Sbf^{ - 1}}{\Xbf_ \bot }\Xbf_ \bot ^H\Sbf_ + ^{ - 1}{\Xbf_{_{//}}}[{\Ibf_M} - {({\Ibf_M} + {\bm\Phi _\Xbf})^{ - 1}}{\bm\Phi _\Xbf}] = \\
{\kern 1pt} {\kern 1pt} {\kern 1pt} {\kern 1pt} {\kern 1pt} {\kern 1pt} {\kern 1pt} {\kern 1pt} {\kern 1pt} {\kern 1pt} {\kern 1pt} {\kern 1pt} {\kern 1pt} {\kern 1pt} {\kern 1pt} {\kern 1pt} {\kern 1pt} {\kern 1pt} {\kern 1pt} {\kern 1pt} {\kern 1pt} {\kern 1pt} {\kern 1pt} {\kern 1pt} {\kern 1pt} {\kern 1pt} {\kern 1pt} {\kern 1pt} {\kern 1pt} {\kern 1pt} {\kern 1pt} {\kern 1pt} {\kern 1pt} {\kern 1pt} {\kern 1pt} {\kern 1pt} {\kern 1pt} {\kern 1pt} {\kern 1pt} {\kern 1pt} {\kern 1pt} {\kern 1pt} {\kern 1pt} {\kern 1pt}  - {\Abf^H}{\Sbf^{ - 1}}{\Xbf_ \bot }\Xbf_ \bot ^H\Sbf_ + ^{ - 1}{\Xbf_{_{//}}}{({\Ibf_M} + {\bm\Phi _\Xbf})^{ - 1}},
\end{array}
\end{equation}
where we have used 
\begin{equation}
\label{48}
	{\Ibf_M} - {({\Ibf_M} + {\bm\Phi _\Xbf})^{ - 1}}{\bm\Phi _\Xbf} = {({\Ibf_M} + {\bm\Phi _\Xbf})^{ - 1}}.
\end{equation}
Substituting \eqref{45} and \eqref{47} into \eqref{44} leas to
\begin{equation}
\label{49}
\Xibf_{\Abf,\Cbf}= {\bm\Psi _{\Abf\Xbf}}{({\Ibf_M} + {\bm\Phi _\Xbf})^{ - 1}},
\end{equation}
where
\begin{equation}
\label{50}
\begin{array}{c}
{\bm\Psi _{\Abf\Xbf}} = {\Abf^H}{\Sbf^{ - 1}}{\Xbf_{_{//}}} - {\Abf^H}{\Sbf^{ - 1}}{\Xbf_ \bot }\Xbf_ \bot ^H\Sbf_ + ^{ - 1}{\Xbf_{_{//}}}\\
 = {\Abf^H}[{\Sbf^{ - 1}} - {\Sbf^{ - 1}}{\Xbf_ \bot }\Xbf_ \bot ^H\Sbf_ + ^{ - 1}]{\Xbf_{_{//}}}.
\end{array}
\end{equation}
According to \eqref{15}, one can verify that
\begin{equation}
\label{51}
\begin{array}{l}
\begin{aligned}
{\Sbf^{ - 1}} - {\Sbf^{ - 1}}{\Sbf_ \bot }\Sbf_ + ^{ - 1} &= {\Sbf^{ - 1}} - {\Sbf^{ - 1}}{\Sbf_ \bot }{({\Sbf_ \bot } + \Sbf)^{ - 1}}\\
 &= {\Sbf^{ - 1}} - {\Sbf^{ - 1}}{\Sbf_ \bot }{[({\Sbf_ \bot }{\Sbf^{ - 1}} + {\Ibf_N})\Sbf]^{ - 1}}\\
 &= {\Sbf^{ - 1}} - {\Sbf^{ - 1}}{\Sbf_ \bot }{\Sbf^{ - 1}}{({\Sbf_ \bot }{\Sbf^{ - 1}} + {\Ibf_N})^{ - 1}}\\
 &= {\Sbf^{ - 1}}[{\Ibf_N} - {\Sbf_ \bot }{\Sbf^{ - 1}}{({\Sbf_ \bot }{\Sbf^{ - 1}} + {\Ibf_N})^{ - 1}}]\\
 &= {\Sbf^{ - 1}}{({\Ibf_N} + {\Sbf_ \bot }{\Sbf^{ - 1}})^{ - 1}}\\
 &= {\Sbf^{ - 1}}{[(\Sbf + {\Sbf_ \bot }){\Sbf^{ - 1}}]^{ - 1}}\\
 &= \Sbf + {\Sbf_ \bot } \\
 & = {\Sbf_ + }.
\end{aligned}
\end{array}
\end{equation}
Inserting \eqref{51} into \eqref{50} results in
\begin{equation}
\label{52}
{\bm\Psi _{\Abf\Xbf}} = {\Abf^H}\Sbf_ + ^{ - 1}{\Xbf_{_{//}}} = {\bm\Phi _{\Abf\Xbf}},
\end{equation}
where ${\bm\Phi _{\Abf\Xbf}}$ is given in \eqref{37}. Substituting \eqref{52} into \eqref{49} leads to
\begin{equation}
\label{53}
\Xibf_{\Abf,\Cbf}= {\bm\Phi _{\Abf\Xbf}}{({\Ibf_M} + {\bm\Phi _\Xbf})^{ - 1}}.
\end{equation}

\subsection{ Equivalent form of \eqref{XiC}}
According to \eqref{32}, we have
\begin{equation}
\label{54}
{({\Ibf_K} + {\Xbf^H}{\Sbf^{ - 1}}\Xbf)^{ - 1}} = {\Ibf_K} - {\Xbf^H}{(\Sbf + \Xbf{\Xbf^H})^{ - 1}}\Xbf.
\end{equation}
Hence,
\begin{equation}
\label{55}
\begin{array}{c}
\begin{aligned}
\Xibf_\Cbf={[{\Ibf_M} - \Xbf_{_{//}}^H{(\Sbf + \Xbf{\Xbf^H})^{ - 1}}{\Xbf_{_{//}}}]^{ - 1}}.
\end{aligned}
\end{array}
\end{equation}
It follows from \eqref{34} that
\begin{equation}
\label{56}
\Xbf_{_{//}}^H{(\Sbf + \Xbf{\Xbf^H})^{ - 1}}{\Xbf_{_{//}}} = {\bm\Psi _\Xbf} - {\bm\Psi _\Xbf}{({\Ibf_M} + {\bm\Psi _\Xbf})^{ - 1}}{\bm\Psi _\Xbf},\end{equation}
where ${\bm\Psi _\Xbf}$ is given in \eqref{39}. Substituting \eqref{56} into \eqref{55} and applying \eqref{46} yields
\begin{equation}
\label{57}
\Xibf_\Cbf = {\Ibf_M} + \Xbf_{_{//}}^H\Sbf_ + ^{ - 1}{\Xbf_{_{//}}}.
\end{equation}

At this point, we can simplify \eqref{31}. According to \eqref{53} and \eqref{57}, the last three multipliers in \eqref{31}, denoted as ${\Mbf_3}$, can be recast as
\begin{equation}
\label{58}
{\Mbf_3} = {\bm\Phi _{\Abf\Xbf}}{({\Ibf_M} + {\bm\Phi _\Xbf})^{ - 1}}{\bm\Phi _{\Xbf\Abf}}.
\end{equation}
In light of \eqref{40} and \eqref{58}, we express \eqref{31} by
\begin{equation}
\label{59}
\begin{array}{l}
\begin{aligned}
{t_\text{GLRGDD}} = {\lambda _{\max }}&\left\{ {\left[ {\bm\Phi _\Abf^{ - 1} + \bm\Phi _\Abf^{ - 1}{\bm\Phi _{\Abf\Xbf}}} \right.} \right.\\
&{\kern 1pt} {\kern 1pt} {\kern 1pt} {\kern 1pt} \left. { \cdot {{({\Ibf_M} + {\bm\Phi _\Xbf} - {\bm\Phi _{\Xbf\Abf}}\bm\Phi _\Abf^{ - 1}{\bm\Phi _{\Abf\Xbf}})}^{ - 1}}{\bm\Phi _{\Xbf\Abf}}\bm\Phi _\Abf^{ - 1}} \right]\\
&{\kern 1pt} {\kern 1pt} {\kern 1pt} {\kern 1pt}  \cdot \left. {{\bm\Phi _{\Abf\Xbf}}{{({\Ibf_M} + {\bm\Phi _\Xbf})}^{ - 1}}{\bm\Phi _{\Xbf\Abf}}} \right\}\\
 = {\lambda _{\max }}&\left\{ {\bm\Phi _\Abf^{ - 1}{\bm\Phi _{\Abf\Xbf}}{{({\Ibf_M} + {\bm\Phi _\Xbf})}^{ - 1}}{\bm\Phi _{\Xbf\Abf}} + \bm\Phi _\Abf^{ - 1}{\bm\Phi _{\Abf\Xbf}}} \right.\\
&{\kern 1pt} {\kern 1pt} {\kern 1pt} {\kern 1pt}  \cdot {({\Ibf_M} + {\bm\Phi _\Xbf} - {\bm\Phi _{\Xbf\Abf}}\bm\Phi _\Abf^{ - 1}{\bm\Phi _{\Abf\Xbf}})^{ - 1}}\\
&{\kern 1pt} {\kern 1pt} {\kern 1pt} {\kern 1pt}  \cdot \left. {{\bm\Phi _{\Xbf\Abf}}\bm\Phi _\Abf^{ - 1}{\bm\Phi _{\Abf\Xbf}}{{({\Ibf_M} + {\bm\Phi _\Xbf})}^{ - 1}}{\bm\Phi _{\Xbf\Abf}}} \right\}.
\end{aligned}
\end{array}
\end{equation}
Using \eqref{29}, we can rewrite \eqref{59} as
\begin{equation}
\label{60}
\begin{array}{l}
\begin{aligned}
{t_\text{GLRGDD}} = {\lambda _{\max }}&\left\{ {{\bm\Phi _{\Xbf\Abf}}\bm\Phi _\Abf^{ - 1}{\bm\Phi _{\Abf\Xbf}}{{({\Ibf_M} + {\bm\Phi _\Xbf})}^{ - 1}}} + {\bm\Phi _{\Xbf\Abf}}\bm\Phi _\Abf^{ - 1}\right.\\
&\cdot{{\bm\Phi _{\Abf\Xbf}}({\Ibf_M} + {\bm\Phi _\Xbf} - {\bm\Phi _{\Xbf\Abf}}\bm\Phi _\Abf^{ - 1}{\bm\Phi _{\Abf\Xbf}})^{ - 1}}\\
&\cdot \left. {{\bm\Phi _{\Xbf\Abf}}\bm\Phi _\Abf^{ - 1}{\bm\Phi _{\Abf\Xbf}}{{({\Ibf_M} + {\bm\Phi _\Xbf})}^{ - 1}}} \right\},
\end{aligned}
\end{array}
\end{equation}
which can be further expressed as
\begin{equation}
\label{61}
\begin{array}{c}
\begin{aligned}
{t_\text{GLRGDD}} =& {\lambda _{\max }}\left\{ {{\bm\Phi _{\Xbf\Abf}}\bm\Phi _\Abf^{ - 1}{\bm\Phi _{\Abf\Xbf}}\left[ {{{({\Ibf_M} + {\bm\Phi _\Xbf})}^{ - 1}}} \right.} \right.\\
{\kern 1pt} {\kern 1pt} {\kern 1pt} {\kern 1pt} & + {({\Ibf_M} + {\bm\Phi _\Xbf} - {\bm\Phi _{\Xbf\Abf}}\bm\Phi _\Abf^{ - 1}{\bm\Phi _{\Abf\Xbf}})^{ - 1}}\\
  &\cdot \left. {\left. {{\bm\Phi _{\Xbf\Abf}}\bm\Phi _\Abf^{ - 1}{\bm\Phi _{\Abf\Xbf}}{{({\Ibf_M} + {\bm\Phi _\Xbf})}^{ - 1}}} \right]} \right\}\\
 =& {\lambda _{\max }}\left\{ {{\bm\Phi _{\Xbf\Abf}}\bm\Phi _\Abf^{ - 1}{\bm\Phi _{\Abf\Xbf}}\left[ {{\Ibf_M}} \right.} \right.\\
{\kern 1pt} {\kern 1pt} {\kern 1pt} {\kern 1pt} & + {({\Ibf_M} + {\bm\Phi _\Xbf} - {\bm\Phi _{\Xbf\Abf}}\bm\Phi _\Abf^{ - 1}{\bm\Phi _{\Abf\Xbf}})^{ - 1}}\\
 &\cdot \left. {\left. {{\bm\Phi _{\Xbf\Abf}}\bm\Phi _\Abf^{ - 1}{\bm\Phi _{\Abf\Xbf}}} \right]{{({\Ibf_M} + {\bm\Phi _\Xbf})}^{ - 1}}} \right\}.
\end{aligned}
\end{array}
\end{equation}
Note that
\begin{equation}
\label{62}
\begin{array}{l}
{({\Ibf_M} + {\bm\Phi _\Xbf} - {\bm\Phi _{\Xbf\Abf}}\bm\Phi _\Abf^{ - 1}{\bm\Phi _{\Abf\Xbf}})^{ - 1}}\\
 {\kern 1pt} {\kern 1pt} {\kern 1pt} {\kern 1pt} {\kern 1pt} {\kern 1pt}  = {\left\{ {({\Ibf_M} + {\bm\Phi _\Xbf})[{\Ibf_M} - {{({\Ibf_M} + {\bm\Phi _\Xbf})}^{ - 1}}{\bm\Phi _{\Xbf\Abf}}\bm\Phi _\Abf^{ - 1}{\bm\Phi _{\Abf\Xbf}}]} \right\}^{ - 1}}\\
 {\kern 1pt} {\kern 1pt} {\kern 1pt} {\kern 1pt} {\kern 1pt}  = {\kern 1pt} {\kern 1pt} {\kern 1pt} {\kern 1pt} {[{\Ibf_M} - {({\Ibf_M} + {\bm\Phi _\Xbf})^{ - 1}}{\bm\Phi _{\Xbf\Abf}}\bm\Phi _\Abf^{ - 1}{\bm\Phi _{\Abf\Xbf}}]^{ - 1}}{({\Ibf_M} + {\bm\Phi _\Xbf})^{ - 1}}{\kern 1pt} {\kern 1pt} {\kern 1pt} .
\end{array}
\end{equation}
Substituting \eqref{62} into \eqref{61} and using \eqref{29} yields
\begin{equation}
\label{63}
\begin{array}{l}
\begin{aligned}
{t_\text{GLRGDD}} = {\lambda _{\max }}&\left( {{{({\Ibf_M} + {\bm\Phi _\Xbf})}^{ - 1}}{\bm\Phi _{\Xbf\Abf}}\bm\Phi _\Abf^{ - 1}{\bm\Phi _{\Abf\Xbf}}} \right.\\
&\cdot\left\{ {{\Ibf_M}}  + {[{\Ibf_M} - {({\Ibf_M} + {\bm\Phi _\Xbf})^{ - 1}}{\bm\Phi _{\Xbf\Abf}}\bm\Phi _\Abf^{ - 1}{\bm\Phi _{\Abf\Xbf}}]^{ - 1}}\right.\\
&\cdot\left. {\left. {{{({\Ibf_M} + {\bm\Phi _\Xbf})}^{ - 1}}{\bm\Phi _{\Xbf\Abf}}\bm\Phi _\Abf^{ - 1}{\bm\Phi _{\Abf\Xbf}}} \right\}} \right).
\end{aligned}
\end{array}
\end{equation}
Define
\begin{equation}
\label{64}
\bm\Theta  = {({\Ibf_M} + {\bm\Phi _\Xbf})^{ - 1}}{\bm\Phi _{\Xbf\Abf}}\bm\Phi _\Abf^{ - 1}{\bm\Phi _{\Abf\Xbf}}.
\end{equation}
Let ${\lambda _{\bm\Theta} }$ be a non-zero eigenvalue of \eqref{64}, and define
\begin{equation}
\label{65}
g({\lambda _{\bm\Theta}  }) = {\lambda _{\bm\Theta}  }[1 + {(1 - {\lambda _{\bm\Theta}  })^{ - 1}}{\lambda _{\bm\Theta}  }].
\end{equation}
According to the spectral norm theorem \cite{LiuXie13b}, $g({\lambda _{\bm\Theta}  })$ is a non-zero eigenvalue of the term within the external parentheses in \eqref{63}. Note that \eqref{65} can be recast as
\begin{equation}
\label{66}
g({\lambda _{\bm\Theta} }) = {(\lambda _{\bm\Theta}  ^{ - 1} - 1)^{ - 1}},
\end{equation}
which can be taken as a monotonically increasing function of ${\lambda _{\bm\Theta} }$. It follows that the GLRGDD in \eqref{63} is equivalent to
\begin{equation}
\label{67}
t_\text{GLRGDD}^{'} = {\lambda _{\max }}\left\{ {{\bm\Phi _{\Xbf\Abf}}\bm\Phi _\Abf^{ - 1}{\bm\Phi _{\Abf\Xbf}}{{({\Ibf_M} + {\bm\Phi _\Xbf})}^{ - 1}}} \right\},
\end{equation}
where we have used \eqref{29}. Substituting \eqref{36} -- \eqref{39} into \eqref{67} results in \eqref{12}. Hence, we can conclude that when $L \ge N$ the GLRGDD--RU and GLRGDD are equivalent.

{\small
\section*{References}
\bibliographystyle{IEEEtran}
\bibliography{D:/LaTexReference/Detection}

\begin{thebibliography}{10}
\providecommand{\url}[1]{#1}
\csname url@samestyle\endcsname
\providecommand{\newblock}{\relax}
\providecommand{\bibinfo}[2]{#2}
\providecommand{\BIBentrySTDinterwordspacing}{\spaceskip=0pt\relax}
\providecommand{\BIBentryALTinterwordstretchfactor}{4}
\providecommand{\BIBentryALTinterwordspacing}{\spaceskip=\fontdimen2\font plus
\BIBentryALTinterwordstretchfactor\fontdimen3\font minus
  \fontdimen4\font\relax}
\providecommand{\BIBforeignlanguage}[2]{{%
\expandafter\ifx\csname l@#1\endcsname\relax
\typeout{** WARNING: IEEEtran.bst: No hyphenation pattern has been}%
\typeout{** loaded for the language `#1'. Using the pattern for}%
\typeout{** the default language instead.}%
\else
\language=\csname l@#1\endcsname
\fi
#2}}
\providecommand{\BIBdecl}{\relax}
\BIBdecl

\bibitem{DeMaioGreco16Book}
A.~De~Maio and S.~G. Greco, \emph{Modern Radar Detection Theory}.\hskip 1em
  plus 0.5em minus 0.4em\relax SciTech Publishing, 2016.

\bibitem{DeMaioOrlando16TSP}
A.~De~Maio, D.~Orlando, C.~Hao, and G.~Foglia, ``Adaptive detection of
  point-like targets in spectrally symmetric interference,'' \emph{IEEE
  Transactions on Signal Processing}, vol.~64, no.~12, pp. 3207--3220, 2016.

\bibitem{XueXu18SP_Knowledge}
J.~Xue, S.~W. Xu, and P.~Shui, ``Knowledge-based adaptive detection of radar
  targets in generalized {Pareto} clutter,'' \emph{Signal Processing}, vol.
  143, pp. 106--111, 2018.

\bibitem{LiuLiu18TSPPHE}
W.~Liu, J.~Liu, Q.~Du, and Y.~L. Wang, ``Distributed target detection in
  partially homogeneous environment when signal mismatch occurs,'' \emph{IEEE
  Transactions on Signal Processing}, vol.~66, no.~14, pp. 3918--3928, 2018.

\bibitem{ShangLiu18GER}
Z.~Shang, Q.~Du, Z.~Tang, T.~Zhang, and W.~Liu, ``Multichannel adaptive signal
  detection in structural nonhomogeneous environment characterized by the
  generalized eigenrelation,'' \emph{Signal Processing}, vol. 148, pp.
  214--222, 2018.

\bibitem{WangZhao19b}
Z.~Wang, Z.~Zhao, C.~Ren, and Z.~Nie, ``{CFAR} subspace detectors with multiple
  observations in system-dependent clutter background,'' \emph{Signal
  Processing}, vol. 153, pp. 58--70, 2018.

\bibitem{LiYang19SP}
N.~Li, H.~Yang, G.~Cui, L.~Kong, and Q.~H. Liu, ``Adaptive two-step {Bayesian}
  {MIMO} detectors in compound-{Gaussian} clutter,'' \emph{Signal Processing},
  vol. 161, pp. 1--13, 2019.

\bibitem{Besson19}
O.~Besson, A.~Coluccia, E.~Chaumette, G.~Ricci, and F.~Vincent, ``Detection of
  {Gaussian} signal using adaptively whitened data,'' \emph{IEEE Signal
  Processing Letters}, vol.~26, no.~3, pp. 430--434, 2019.

\bibitem{AubryDeMaio20TSP}
A.~{Aubry}, A.~{De Maio}, S.~{Marano}, and M.~{Rosamilia}, ``Single-pulse
  simultaneous target detection and angle estimation in a multichannel phased
  array radar,'' \emph{IEEE Transactions on Signal Processing}, vol.~68, pp.
  6649--6664, 2020.

\bibitem{Kelly86}
E.~J. Kelly, ``An adaptive detection algorithm,'' \emph{IEEE Transactions on
  Aerospace and Electronic Systems}, vol.~22, no.~1, pp. 115--127, 1986.

\bibitem{RobeyFuhrmann92}
F.~C. Robey, D.~R. Fuhrmann, E.~J. Kelly, and R.~Nitzberg, ``A {CFAR} adaptive
  matched filter detector,'' \emph{IEEE Transactions on Aerospace and
  Electronic Systems}, vol.~28, no.~1, pp. 208--216, 1992.

\bibitem{DeMaio07}
A.~De~Maio, ``{Rao} test for adaptive detection in {Gaussian} interference with
  unknown covariance matrix,'' \emph{IEEE Transactions on Signal Processing},
  vol.~55, no.~7, pp. 3577--3584, 2007.

\bibitem{DeMaio04}
------, ``A new derivation of the adaptive matched filter,'' \emph{IEEE Signal
  Processing Letters}, vol.~11, no.~10, pp. 792--793, 2004.

\bibitem{KrautScharf99}
S.~Kraut and L.~L. Scharf, ``The {CFAR} adaptive subspace detector is a
  scale-invariant {GLRT},'' \emph{IEEE Transactions on Signal Processing},
  vol.~47, no.~9, pp. 2538--2541, 1999.

\bibitem{DeMaio05}
A.~De~Maio, ``Robust adaptive radar detection in the presence of steering
  vector mismatches,'' \emph{IEEE Transactions on Aerospace and Electronic
  Systems}, vol.~41, no.~4, pp. 1322--1337, 2005.

\bibitem{DeMaioDeNicola09b}
A.~De~Maio, S.~De~Nicola, Y.~Huang, S.~Zhang, and A.~Farina, ``Adaptive
  detection and estimation in the presence of useful signal and interference
  mismatches,'' \emph{IEEE Transactions on Signal Processing}, vol.~57, no.~2,
  pp. 436--450, 2009.

\bibitem{HaoOrlando14a}
C.~Hao, D.~Orlando, X.~Ma, S.~Yan, and C.~Hou, ``Persymmetric detectors with
  enhanced rejection capabilities,'' \emph{IET Radar, Sonar and Navigation},
  vol.~8, no.~5, pp. 557--563, 2014.

\bibitem{PulsoneRader01}
N.~B. Pulsone and C.~M. Rader, ``Adaptive beamformer orthogonal rejection
  test,'' \emph{IEEE Transactions on Signal Processing}, vol.~49, no.~3, pp.
  521--529, 2001.

\bibitem{BandieraBesson07TSP_WABORT}
F.~Bandiera, O.~Besson, and G.~Ricci, ``An {ABORT}-like detector with improved
  mismatched signals rejection capabilities,'' \emph{IEEE Transactions on
  Signal Processing}, vol.~56, no.~1, pp. 14--25, 2007.

\bibitem{OrlandoRicci10}
D.~Orlando and G.~Ricci, ``A {Rao} test with enhanced selectivity properties in
  homogeneous scenarios,'' \emph{IEEE Transactions on Signal Processing},
  vol.~58, no.~10, pp. 5385--5390, 2010.

\bibitem{LiuLiu21SCIS}
W.~Liu, J.~Liu, C.~Hao, Y.~Gao, and Y.-L. Wang, ``Multichannel adaptive signal
  detection: Basic theory and literature review,'' \emph{Science China:
  Information Sciences}, DOI: 10.1007/s11432-020-3211-8, 2021.

\bibitem{BessonTourneret07}
O.~Besson, J.-Y. Tourneret, and S.~Bidon, ``Knowledge-aided {Bayesian}
  detection in heterogeneous environments,'' \emph{IEEE Signal Processing
  Letters}, vol.~14, no.~5, pp. 355--358, 2007.

\bibitem{BidonBesson11AES}
S.~Bidon, O.~Besson, and J.-Y. Tourneret, ``Knowledge-aided {STAP} in
  heterogeneous clutter using a hierarchical {Bayesian} algorithm,'' \emph{IEEE
  Transactions on Aerospace and Electronic Systems}, vol.~47, no.~3, pp.
  1863--1879, 2011.

\bibitem{StoicaLiZhu08}
P.~Stoica, J.~Li, X.~Zhu, and J.~R. Guerci, ``On using a priori knowledge in
  space-time adaptive processing,'' \emph{IEEE Transactions on Signal
  Processing}, vol.~56, no.~6, pp. 2598--2602, 2008.

\bibitem{DeMaioFarina10}
A.~De~Maio, A.~Farina, and G.~Foglia, ``Knowledge-aided {Bayesian} radar
  detectors {\&} their application to live data,'' \emph{IEEE Transactions on
  Aerospace and Electronic Systems}, vol.~46, no.~1, pp. 170--183, 2010.

\bibitem{BandieraBesson11}
F.~Bandiera, O.~Besson, and G.~Ricci, ``Adaptive detection of distributed
  targets in compound-{Gaussian} noise without secondary data: A {Bayesian}
  approach,'' \emph{IEEE Transactions on Signal Processing}, vol.~59, no.~12,
  pp. 5698--5708, 2011.

\bibitem{ConteDeMaio01}
E.~Conte, A.~De~Maio, and G.~Ricci, ``{GLRT}-based adaptive detection
  algorithms for range-spread targets,'' \emph{IEEE Transactions on Signal
  Processing}, vol.~49, no.~7, pp. 1336--1348, 2001.

\bibitem{BessonScharf06b}
O.~Besson, L.~L. Scharf, and S.~Kraut, ``Adaptive detection of a signal known
  only to lie on a line in a known subspace, when primary and secondary data
  are partially homogeneous,'' \emph{IEEE Transactions on Signal Processing},
  vol.~54, no.~12, pp. 4698--4705, 2006.

\bibitem{LiuLiu2015b}
W.~Liu, J.~Liu, L.~Huang, K.~Yan, and Y.~Wang, ``Robust {GLRT} approaches to
  signal detection in the presence of spatial-temporal uncertainty,''
  \emph{Signal Processing}, vol. 118, pp. 272--284, 2016.

\bibitem{GauReed98}
Y.-L. Gau and I.~S. Reed, ``An improved reduced-rank {CFAR} space-time adaptive
  radar detection algorithm,'' \emph{IEEE Transactions on Signal Processing},
  vol.~46, no.~8, pp. 2139--2146, 1998.

\bibitem{AyoubHaimovich00}
T.~F. Ayoub and A.~M. Haimovich, ``Modified {GLRT} signal detection
  algorithm,'' \emph{IEEE Transactions on Aerospace and Electronic Systems},
  vol.~36, no.~3, pp. 810--818, 2000.

\bibitem{JinFriedlander05b}
Y.~Jin and B.~Friedlander, ``Reduced-rank adaptive detection of distributed
  sources using subarrays,'' \emph{IEEE Transactions on Signal Processing},
  vol.~53, no.~1, pp. 13--25, 2005.

\bibitem{AubryDeMaio13}
A.~Aubry, A.~De~Maio, L.~Pallotta, and A.~Farina, ``Radar detection of
  distributed targets in homogeneous interference whose inverse covariance
  structure is defined via unitary invariant functions,'' \emph{IEEE
  Transactions on Signal Processing}, vol.~61, no.~20, pp. 4949--4961, 2013.

\bibitem{WangSahinoglu11}
P.~Wang, Z.~Sahinoglu, M.-O. Pun, H.~Li, and B.~Himed, ``Knowledge-aided
  adaptive coherence estimator in stochastic partially homogeneous
  environments,'' \emph{IEEE Signal Processing Letters}, vol.~18, no.~3, pp.
  193--196, 2011.

\bibitem{ZhouZhang12}
Y.~Zhou and L.-R. Zhang, ``Knowledge-aided {Bayesian} radar adaptive detection
  in heterogeneous environment: {GLRT}, {Rao} and {Wald} test,''
  \emph{International Journal of Electronics and Communications (AEU)},
  vol.~66, no.~3, pp. 239--243, 2012.

\bibitem{LiuHan18}
J.~{Liu}, J.~{Han}, Z.~{Zhang}, and J.~{Li}, ``Bayesian detection for {MIMO}
  radar in {Gaussian} clutter,'' \emph{IEEE Transactions on Signal Processing},
  vol.~66, no.~24, pp. 6549--6559, Dec 2018.

\bibitem{HaoMa12}
C.~Hao, X.~Ma, X.~Shang, and L.~Cai, ``Adaptive detection of distributed
  targets in partially homogeneous environment with {Rao} and {Wald} tests,''
  \emph{Signal Processing}, vol.~92, no.~4, pp. 926--930, 2012.

\bibitem{HaoOrlando14b}
C.~Hao, D.~Orlando, G.~Foglia, X.~Ma, S.~Yan, and C.~Hou, ``Persymmetric
  adaptive detection of distributed targets in partially-homogeneous
  environment,'' \emph{Digital Signal Processing}, vol.~24, pp. 42--51, 2014.

\bibitem{GaoLiao14}
Y.~Gao, G.~Liao, S.~Zhu, X.~Zhang, and D.~Yang, ``Persymmetric adaptive
  detectors in homogeneous and partially homogeneous environments,'' \emph{IEEE
  Transactions on Signal Processing}, vol.~62, no.~2, pp. 331--342, 2014.

\bibitem{LiuLiu18TSPPsmtrc}
J.~{Liu}, W.~{Liu}, Y.~{Gao}, S.~{Zhou}, and X.~{Xia}, ``Persymmetric adaptive
  detection of subspace signals: Algorithms and performance analysis,''
  \emph{IEEE Transactions on Signal Processing}, vol.~66, no.~23, pp.
  6124--6136, Dec 2018.

\bibitem{LiuSun19}
J.~Liu, S.~Sun, and W.~Liu, ``One-step persymmetric {GLRT} for subspace
  signals,'' \emph{IEEE Transaction on Signal Processing}, vol.~14, no.~67, pp.
  3639--3648, July 15 2019.

\bibitem{BoseSteinhardt96a}
S.~Bose and A.~O. Steinhardt, ``Adaptive array detection of uncertain rank one
  waveforms,'' \emph{IEEE Transactions on Signal Processing}, vol.~44, no.~11,
  pp. 2801--2809, 1996.

\bibitem{LiuXie14c}
W.~Liu, W.~Xie, J.~Liu, D.~Zou, H.~Wang, and Y.~Wang, ``Detection of a
  distributed target with direction uncertainty,'' \emph{IET Radar, Sonar and
  Navigation}, vol.~8, no.~9, pp. 1177--1183, 2014.

\bibitem{LiuZhang15}
W.~{Liu}, W.~{Zhang}, H.~{Li}, C.~{Zhang}, Y.~{Wang}, and J.~{Liu},
  ``Generalized direction detectors in sample-starved environments,'' in
  \emph{2015 IEEE China Summit and International Conference on Signal and
  Information Processing (ChinaSIP)}, July 2015, pp. 296--299.

\bibitem{LiuLi19b}
J.~{Liu} and J.~{Li}, ``Robust detection in {MIMO} radar with steering vector
  mismatches,'' \emph{IEEE Transactions on Signal Processing}, vol.~67, no.~20,
  pp. 5270--5280, Oct 2019.

\bibitem{LiuLiu20a}
W.~Liu, J.~Liu, Y.~Gao, G.~Wang, and Y.-L. Wang, ``Multichannel signal
  detection in interference and noise when signal mismatch happens,''
  \emph{Signal Processing}, vol. 166, p. 107268, 2020.

\bibitem{WicksKASTAP}
M.~C. Wicks, M.~Rangaswamy, R.~Adve, and T.~B. Hale, ``Space-time adaptive
  processing: a knowledge-based perspective for airborne radar,'' \emph{IEEE
  Signal Processing Magazine}, vol.~23, no.~1, pp. 51--65, 2006.

\bibitem{LiuXie14b}
W.~Liu, W.~Xie, J.~Liu, and Y.~Wang, ``Adaptive double subspace signal
  detection in {Gaussian} background--part {I}: Homogeneous environments,''
  \emph{IEEE Transactions on Signal Processing}, vol.~62, no.~9, pp.
  2345--2357, 2014.

\bibitem{Anderson03}
T.~W. Anderson, \emph{An Introduction to Multivariate Statistical Analysis},
  3rd~ed.\hskip 1em plus 0.5em minus 0.4em\relax Hoboken: Wiley, 2003.

\bibitem{BandieraBesson07}
F.~Bandiera, O.~Besson, D.~Orlando, G.~Ricci, and L.~L. Scharf, ``{GLRT}-based
  direction detectors in homogeneous noise and subspace interference,''
  \emph{IEEE Transactions on Signal Processing}, vol.~55, no.~6, pp.
  2386--2394, 2007.

\bibitem{Kay98}
S.~M. Kay, \emph{Fundamentals of Statistical Signal Processing: Detection
  Theory}.\hskip 1em plus 0.5em minus 0.4em\relax Englewood Cliffs, NJ:
  Prentice-Hall, 1998.

\bibitem{LiuXie13b}
W.~Liu, W.~Xie, and Y.~Wang, ``{Rao} and {Wald} tests for distributed targets
  detection with unknown signal steering,'' \emph{IEEE Signal Processing
  Letters}, vol.~20, no.~11, pp. 1086--1089, 2013.

\end{thebibliography}
}

\begin{figure}[htbp]
\setlength{\abovecaptionskip}{2pt}
\setlength{\belowcaptionskip}{2pt}
\centering
\includegraphics[width=1.0\textwidth]{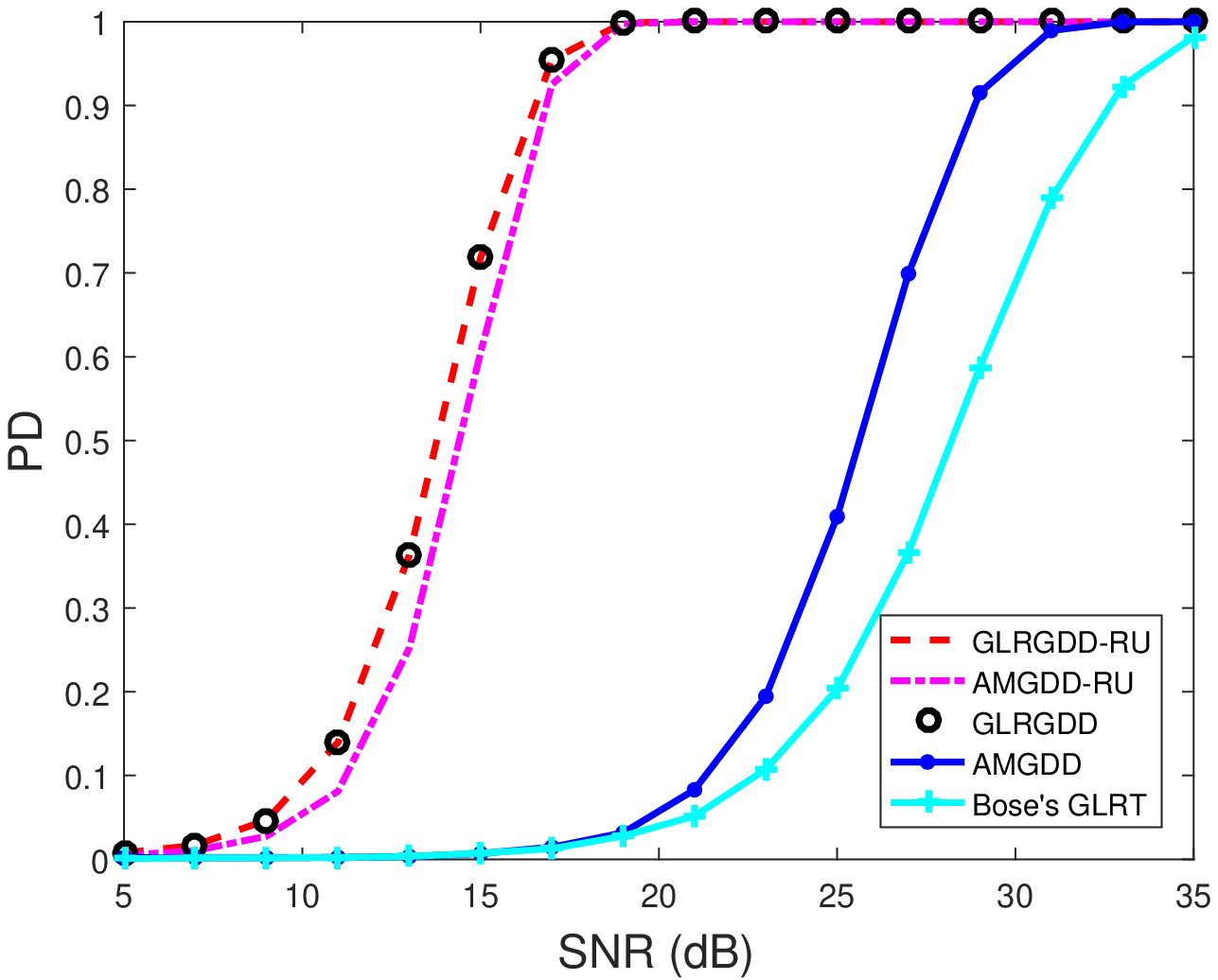} 
\\
{Fig.1.~~ PD versus SNR. $N = 12$, $J = 2$, $M = 3$, $K = 16$, and $L = 14$.}
\label{PD_SNR_N12_J2_M3_K16_L14}
\end{figure}

\begin{figure}[htbp]
\setlength{\abovecaptionskip}{2pt}
\setlength{\belowcaptionskip}{2pt}
\centering
\includegraphics[width=1.0\textwidth]{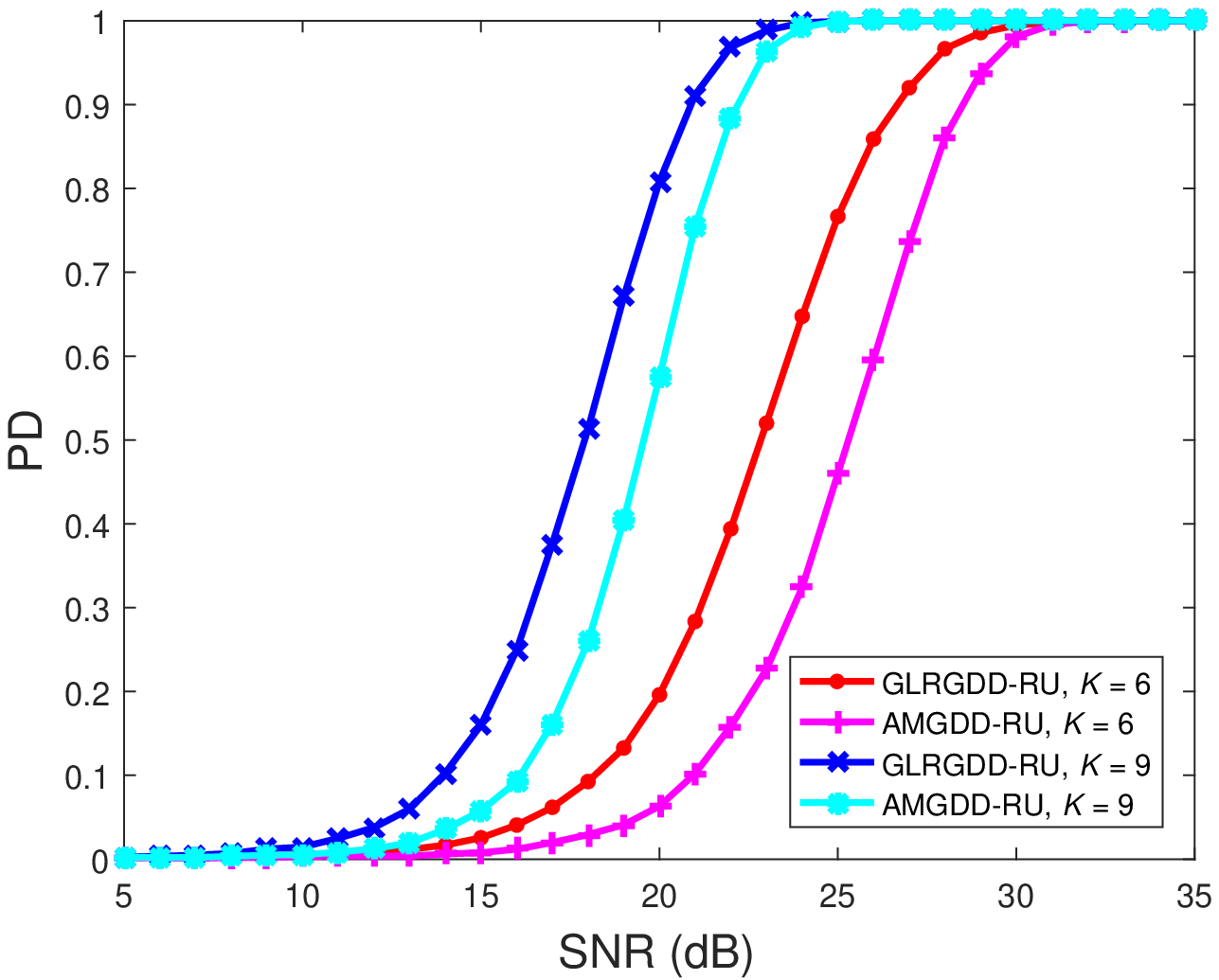} 
\\
{Fig.2.~~ PD versus SNR. $N = 12$, $J = 2$, $M = 3$, and $L = 11$.}
\label{PD_SNR_N12_J2_M3_L11}
\end{figure}

%

\end{document}